\documentclass{article}

\usepackage{arxiv}

\usepackage[utf8]{inputenc} 
\usepackage[T1]{fontenc}    
\usepackage{hyperref}       
\usepackage{url}            
\usepackage{booktabs}       
\usepackage{amsfonts}       
\usepackage{nicefrac}       
\usepackage{microtype}      
\usepackage{lipsum}
\usepackage{graphicx}
\usepackage{multirow}
\usepackage{colortbl}
\graphicspath{ {./images/} }

\title{On the importance of learning non-local dynamics for stable data-driven climate modeling: A 1D gravity wave-QBO testbed}

\author{
 Hamid A. Pahlavan \\
  NorthWest Research Associates \\
  Boulder, CO, USA\\
  \texttt{pahlavan@nwra.com}
   \And
 Pedram Hassanzadeh \\
    University of Chicago \\
    Chicago, IL, USA \\
    \texttt{pedramh@uchicago.edu} \\
  \And
 M. Joan Alexander \\
  NorthWest Research Associates \\
  Boulder, CO, USA\\
  \texttt{alexand@nwra.com} \\
}

\begin{document}
\maketitle
\begin{abstract}
Machine learning (ML) techniques, especially neural networks (NNs), have shown promise in learning subgrid-scale parameterizations for climate models. However, a major problem with data-driven parameterizations, particularly those learned with supervised algorithms, is model instability. Current remedies are often ad-hoc and lack a theoretical foundation. Here, we combine ML theory and climate physics to address a source of instability in NN-based parameterization. We demonstrate the importance of learning spatially \textit{non-local} dynamics using a 1D model of the quasi-biennial oscillation (QBO) with gravity wave (GW) parameterization as a testbed. While common offline metrics fail to identify shortcomings in learning non-local dynamics, we show that the concept of receptive field (RF) can identify instability a-priori. We find that NN-based parameterizations that seem to accurately predict GW forcings from wind profiles ($\mathbf{R^2 \approx 0.99}$) cause unstable simulations when RF is too small to capture the non-local dynamics, while NNs of the same size but large-enough RF are stable. We examine three broad classes of architectures, namely convolutional NNs, Fourier neural operators, and fully-connected NNs; the latter two have inherently large RFs. We also demonstrate that learning non-local dynamics is crucial for the stability and accuracy of a data-driven spatiotemporal emulator of the zonal wind field. Given the ubiquity of non-local dynamics in the climate system, we expect the use of effective RF, which can be computed for any NN architecture, to be important for many applications. This work highlights the necessity of integrating ML theory with physics to design and analyze data-driven algorithms for weather and climate modeling.

\end{abstract}



\section*{Introduction}
Accurate climate projections are of significant societal and scientific importance. These projections currently depend on general circulation models (GCMs) with grid spacings of tens to hundreds of kilometers, which cannot resolve smaller subgrid-scale (SGS) processes. Since SGS processes significantly influence Earth's climate, they must be represented using parameterizations. Traditional parameterizations, which rely on simplified physical models and heuristic approximations, suffer from inaccuracies that lead to significant uncertainties \cite{schneider2017climate}. A promising alternative is to use machine learning (ML) techniques, particularly neural networks (NNs), to develop novel data-driven parameterizations with reduced parametric and structural uncertainties \cite{o2018using, gentine2018could, bolton2019applications, yuval2020stable, zanna2020data, han2020moist, guan2023learning, jakhar2023learning,lai2024machine}. However, data-driven parameterizations, particularly those learned offline using supervised algorithms, often lead to instabilities and drifts when implemented online in GCMs \cite{rasp2018deep, brenowitz2018prognostic, brenowitz2019spatially, brenowitz2020interpreting, rasp2020coupled, iglesias2024causally}.

The reasons behind these instabilities are often not fully understood, and remedies are typically ad-hoc, lacking a theoretical basis or robust methodology. For instance, instabilities in NN-based parameterizations of atmospheric SGS tendencies (e.g., heating and moistening due to convection) have been addressed by using deeper architectures and intensive hyperparameter tuning \cite{rasp2018deep} or by removing variables from the NN's input that might lead to non-causal relationships between input and output \cite{brenowitz2018prognostic, brenowitz2019spatially}. In parameterizing ocean mesoscale eddies, stability has been achieved by attenuating the strength of the parameterization feedback \cite{zanna2020data, jakhar2023learning}. Adding to the challenge, there is often a disconnect between offline metrics and online performance, and success in offline settings does not always translate to online performance \cite{Brenowitz2020-do, lin2023systematic}. However, some physics-based analyses have shown potential for identifying the reasons behind instability and the discrepancy between offline and online performance, though they are problem-specific  \cite{pahlavan2024explainable, guan2022stable, brenowitz2020interpreting}.

Such instabilities are not limited to NN-based parameterizations; they can also occur in ocean or atmosphere `emulators' \cite{chattopadhyay2023long, subel2024building}, such as NN-based weather forecasting models, where NNs autoregressively predict a future state from a previous state \cite{pathak2022fourcastnet,bi2023accurate, lam2023learning}. To date, rigorous frameworks for analyzing, understanding, and systematically rectifying such instabilities remain elusive.

In this study, we use a conceptual one-dimensional (1D) model of the quasi-biennial oscillation (QBO) and gravity wave (GW) drag parameterizations \cite{holton1972updated, plumb1977interaction} (\textit{Materials and Methods A}) as a testbed to demonstrate one source of such instabilities and present rigorous methods for addressing them. The QBO is a quasi-periodic fluctuation between eastward and westward zonal mean zonal wind, with an average period of $\sim$28 months \cite{baldwin2001quasi}, and is the primary mode of interannual variability in the tropical stratosphere with links to subseasonal-to-seasonal forecast skills \cite{anstey2022impacts}. Despite its simplicity, the 1D model effectively captures the essentials of wave-mean flow interaction, a fundamental aspect of stratospheric dynamics. Therefore, the 1D model has been extensively used to investigate various dynamics of the QBO and its wave forcing \cite{renaud2019periodicity, leard2020multimodal, match2020mean, match2021anomalous, chartrand2024recovering}, as well as to advance the development of ML-based GW parameterizations \cite{hardiman2023machine, pahlavan2024explainable, shamir2024graft}.
Here, we aim to learn the GW drag in the 1D model (\(G\) in Eq.~\ref{eq:1}). Hence, we offline train NN-based GW parameterizations that receive resolved wind profiles (\(u\) in Eq.~\ref{eq:1}) as input and predict GW drag (an SGS term). Within this context, we:

\begin{itemize}

\item Demonstrate that NNs of the same size and similar offline accuracy (based on pattern correlation and root-mean-square-error (RMSE)), can lead to either stable or unstable simulations, depending on their ability to learn the crucial non-local dynamics of GW-mean flow interactions.

\item Introduce a metric based on ML theory, the (effective) receptive field (RF), which can be calculated for any NN architecture and quantify the degree to which non-locality is captured. We show that RF can predict the instability of NN-based GW parameterizations \textit{a priori} as an offline metric.
\end{itemize}
Furthermore, based on this insight from SGS modeling, we
\begin{itemize}
\item Demonstrate that failing to capture non-local dynamics can also lead to an unstable spatiotemporal emulator (which autoregressively predicts the evolution of the full state \(u(t)\)).
\end{itemize}

We will consider three different NN architectures, representing three broad classes of algorithms; however, our insight and approach apply to any NN architecture. Our primary focus is on convolutional neural networks (CNNs), which extract hierarchical features from input data using convolutional filters \cite{goodfellow2016deep}. CNNs are extensively used in climate science, including for SGS parameterizations \cite{bolton2019applications, guillaumin2021stochastic, frezat2021physical, guan2022stable, liu2022investigation, subel2023explaining, guan2023learning, hardiman2023machine, qu2023can, srinivasan2024turbulence}, due to their exceptional pattern extraction capabilities. Additionally, we use the Fourier neural operator (FNO) \cite{li2020fourier}, known for its efficiency in learning non-local dynamics \cite{lanthaler2023nonlocal}. The FNO leverages the Fourier transform to learn mappings between function spaces, capturing complex spatial patterns and dynamics in high-dimensional data \cite{li2020fourier}. Finally, we use a multi-layer perceptron (MLP), a class of NNs composed of multiple layers of neurons, where each neuron in a layer is connected to every neuron in the next layer. Because of this dense connectivity, MLPs tend to require more parameters and are generally less efficient. More details regarding these NNs are provided in \textit{Materials and Methods B}.

\section*{Results}
\label{sec:Results}
\begin{figure*}[t!]
\centering
\includegraphics[width=11.4cm]{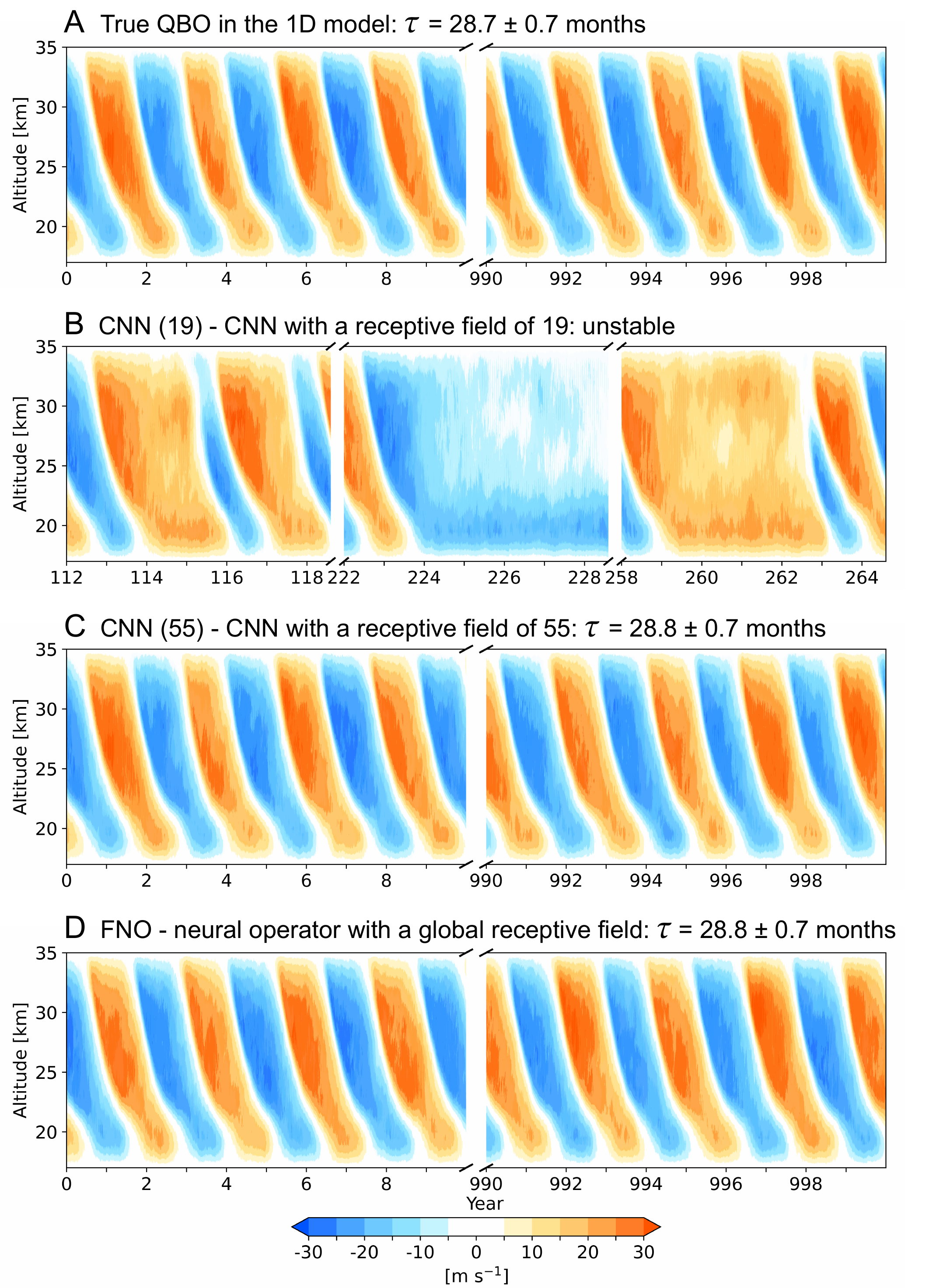}
\caption{Quasi-biennial oscillation (QBO), as seen in a time-height section of zonal wind. (\textit{A}) The ``true" QBO in the 1D model with a mean period ($\tau$) of 28.7 months, and a standard deviation of 0.7 months. Online performance of the different NN architectures that predict GW drag $G$ as a function of zonal wind $u$: (\textit{B} and \textit{C}) CNNs with RF of 19 and 55, respectively, and (\textit{D}) FNO. Each panel displays only 20 years of a 1000-year simulation for illustration.}\label{fig. 1}
\end{figure*}

\paragraph{A. Instability of NN-based parameterizations.} Figure~\ref{fig. 1}\textit{A} shows the QBO in the 1D model that we define here as the ``true QBO": an oscillation between eastward and westward wind with a period ($\tau$) of 28.7$\pm$0.7 months (\textit{Materials and Methods A}). The standard deviation is based on $\sim$430 QBO cycles in a 1000-year simulation. This simulation serves as our ``truth" to evaluate the performance of NN-based GW parameterizations. Using the same setup, we produce an independent 100-year dataset for training and validation; see \cite{pahlavan2024explainable} for more details about the development of NN-based GW parameterizations in this model.

Figure~\ref{fig. 1}\textit{B} shows the online performance of a CNN-based GW parameterization, i.e., after coupling the NN with the 1D model. Despite CNN's excellent offline performance ($R^2 \approx 0.99$ between the true and CNN-predicted $G$), the simulation is unstable and exhibits unrealistic transitions between QBO phases. Three $\sim$7-year periods from a 1000-year simulation are shown in this panel to highlight these unrealistic transitions. Notably, the model generates QBO cycles where either the eastward or westward phases persist for excessively long or short durations. Figure~\ref{fig. 1}\textit{C} displays the online performance of another CNN with similarly great offline performance. This CNN has the same number of parameters (approximately 15,000) and layers (four) as the one in Fig.~\ref{fig. 1}\textit{B}, but uses a significantly larger kernel size (19 vs. 7).  The simulation shows remarkable online accuracy and stability, with realistic QBO phase transitions throughout the 1000-year simulation and a standard deviation of 0.7 months for the period, matching the true QBO. Similarly, an FNO with the same number of parameters and comparable offline performance also demonstrates stable and accurate online performance (Fig.~\ref{fig. 1}\textit{D}). The MLP-based GW parameterization also achieves great online performance (not shown).

We emphasize that the stable and unstable CNNs have the same number of parameters, depth, and offline accuracy based on the commonly used metrics of RMSE and $R^2$. As mentioned earlier, such a disconnect between offline and online performance has been reported in many past studies for various climate processes and is often attributed to the deficiencies of the supervised (offline) learning approach. However, our testbed enables a deeper investigation to understand the sources of the instability and to develop systematic approaches to address them. In particular, the instability shown in Fig.~\ref{fig. 1}\textit{B} seems most pronounced during the transition phases of the QBO, hence our emphasis on the standard deviation of the period as a measure of stability. This suggests that errors at the topmost levels of the 1D model are significant. Given our understanding of the non-local nature of GW propagation and dissipation and the development of this instability, we next examine how well the NNs capture non-locality by calculating their RF, as explained below.


\begin{figure*}[t!]
\centering
\includegraphics[width=17.8cm]{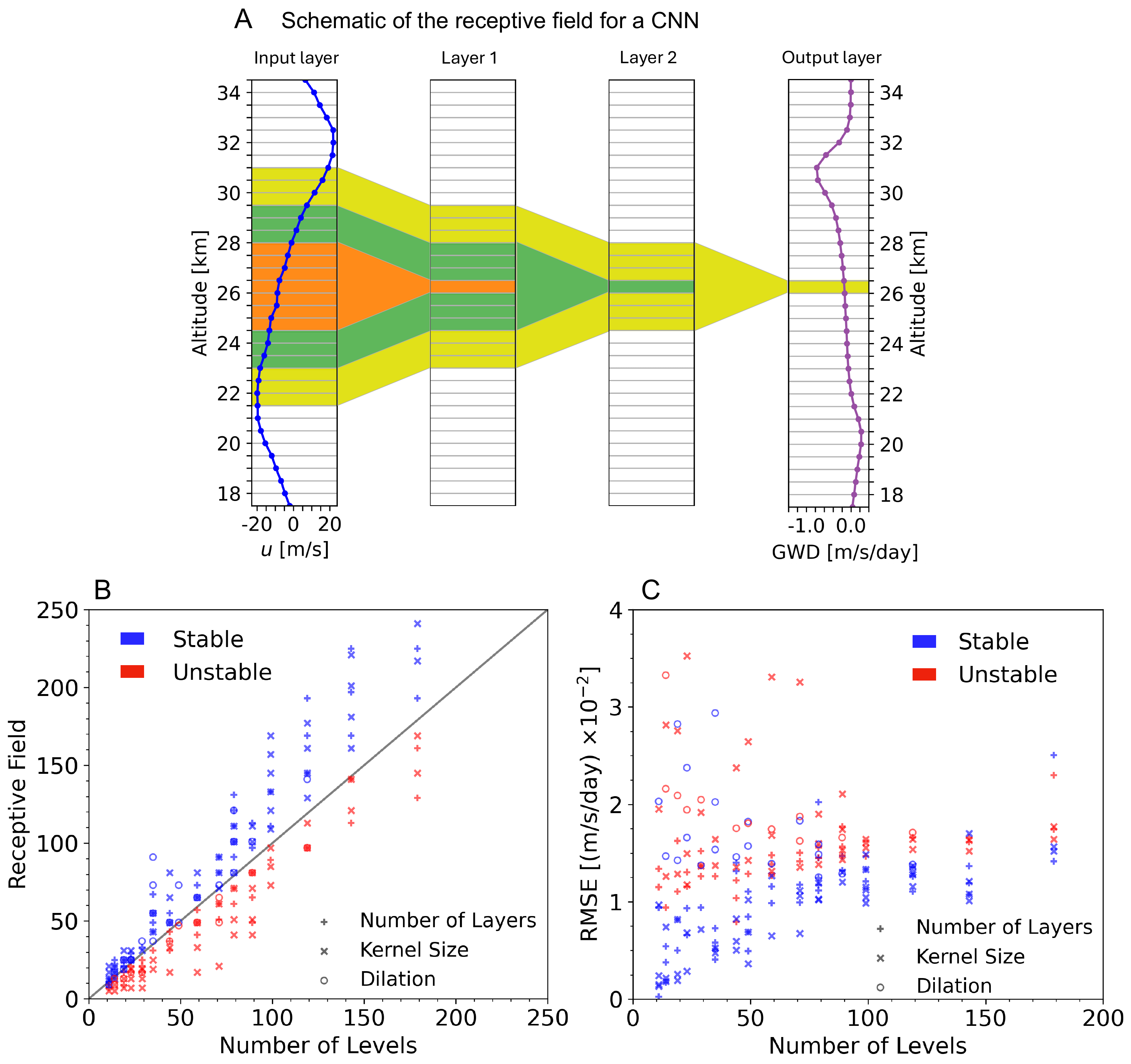}
\caption{The RF of CNNs and its relation to their performance. (\textit{A}) Schematic of the RF for an output in the middle of the vertical domain for various layers of a CNN with a kernel size of seven. The output at the final layer (i.e., GW drag at the 26 km level) can \textit{see} only 19 levels of the input (i.e., zonal wind profile). For a vertical resolution of 500 m, it covers 9.5 km of the input domain, ranging from 21.5 to 31 km. (\textit{B}) Online stability of CNNs with varying RFs for different resolutions (number of levels) of the 1D model. The RF of a CNN can be increased by adding layers, enlarging the kernel size, or expanding the dilation, as indicated. (\textit{C}) Similar to \textit{B}, but showing the RMSE of CNNs instead of RF.}\label{fig. 2}
\end{figure*}

\begin{figure*}[t!]
\centering
\includegraphics[width=11.4cm]{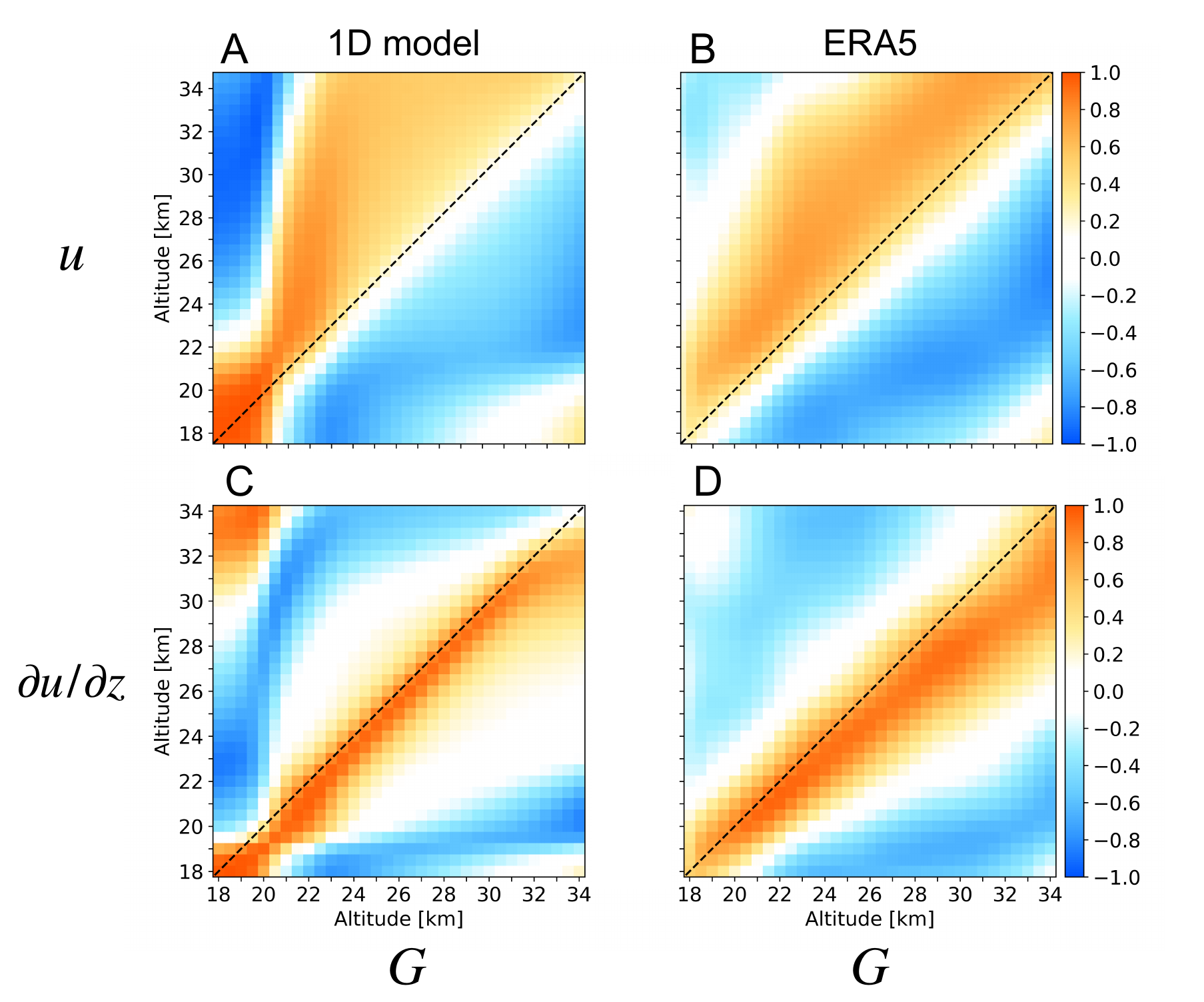}
\caption{Correlation matrices highlight the non-locality of the GW dynamics. (\textit{A} and \textit{B}) Correlation between zonal wind ($u$) and GW drag ($G$), and (\textit{C} and \textit{D}) between the vertical shear of zonal wind ($\partial{u}/\partial{z}$) and $G$ in the 1D model, and the observed QBO based on the ERA5 reanalysis data (2010-2019), as indicated. For ERA5, $u$ is the zonal-mean zonal wind averaged over 5$^{\circ}$N–5$^{\circ}$S, and $G$ is the zonal wave forcing (divergence of the upward flux of zonal momentum). Note that $G$ in ERA5 is due to resolved waves, primarily GWs but also Kelvin waves.}\label{fig. 3}
\end{figure*}

\begin{figure*}[t!]
\centering
\includegraphics[width=11.4cm]{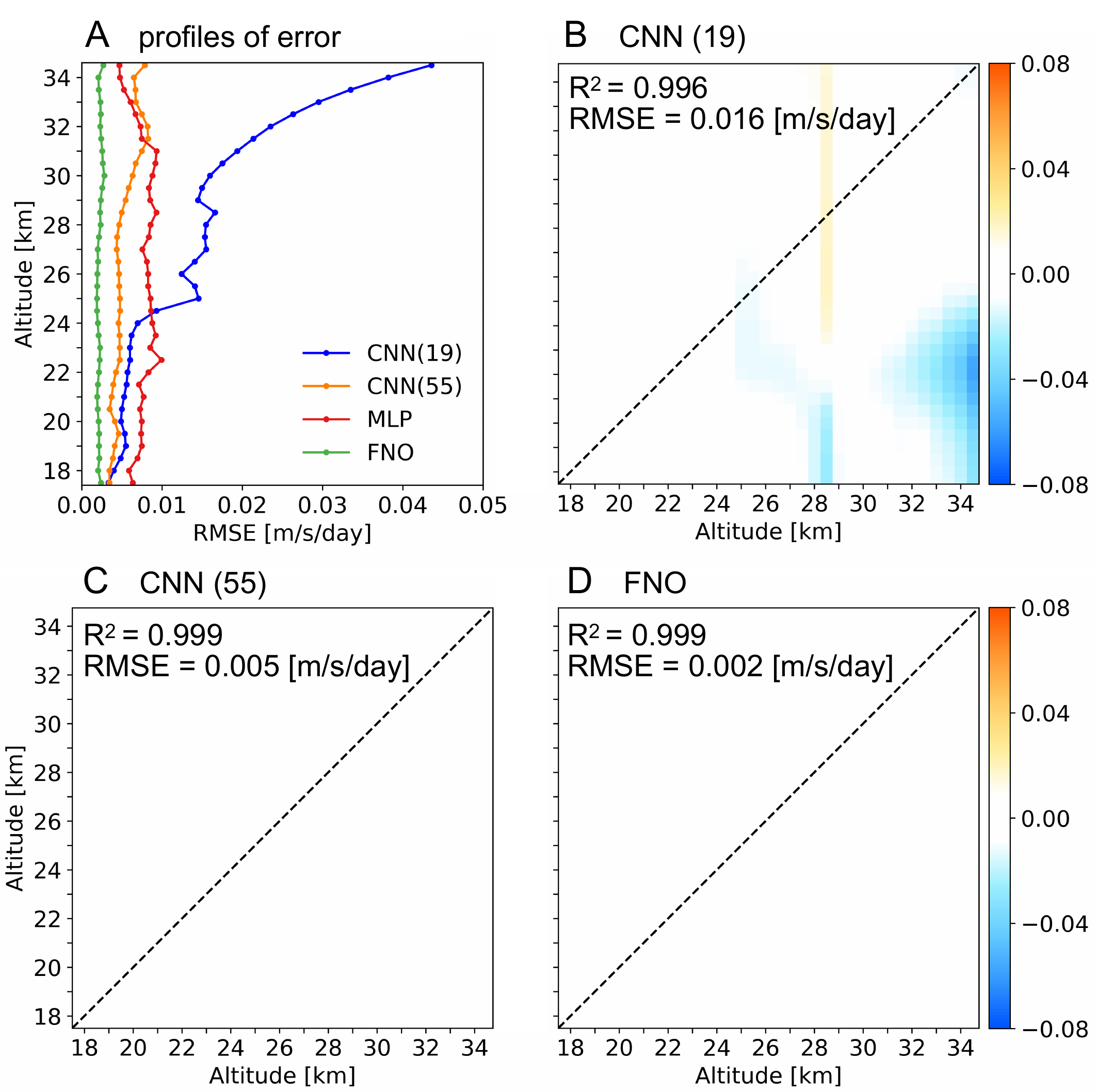}
\caption{Offline error of various NN architectures. (\textit{A}) Vertical profile of error for different NN architectures as indicated. (\textit{B}-\textit{D}) Difference between the true and NN-predicted correlation matrices for zonal wind and GW drag, using CNN(19), CNN(55), and FNO, respectively. This difference is the NN-predicted correlation matrix (not shown) subtracted from the true correlation matrix shown in Fig.~\ref{fig. 3}\textit{A}. The result for the MLP (\(R^2=0.999\), RMSE=0.007 [m/s/day]) is not shown, but it is similar to (\textit{C} and \textit{D}) showing no significant difference from the true correlation matrix.}\label{fig. 4}
\end{figure*}

\begin{figure*}[t!]
\centering
\includegraphics[width=11.4cm]{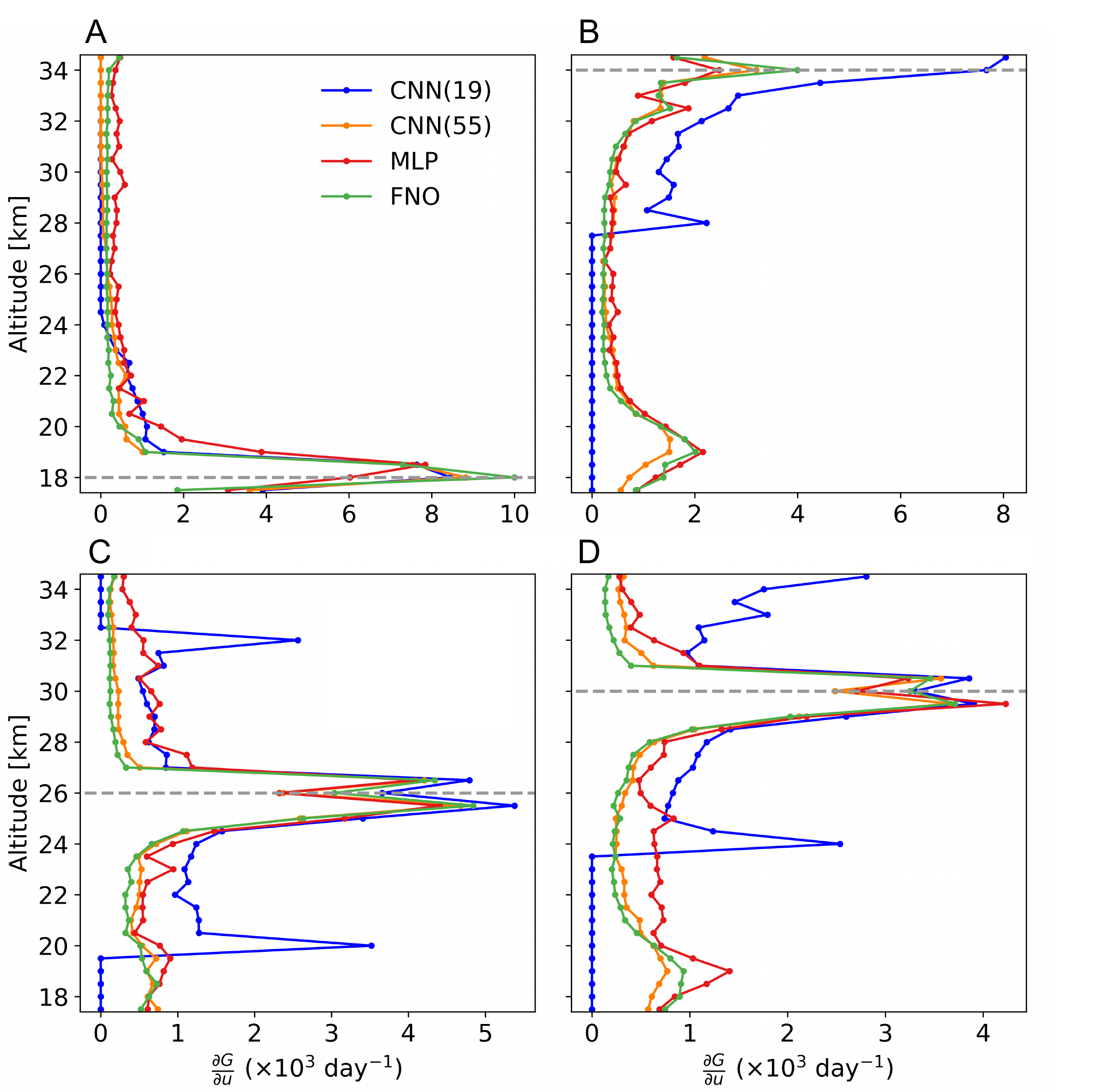}
\caption{Effective receptive field (ERF) for GW drag at levels (\textit{A}) 18 km, (\textit{B}) 34 km, (\textit{C}) 26 km, and (\textit{D}) 30 km (marked by dashed horizontal grey lines) for various NN architectures, as indicated. The ERF is calculated by back-propagating an arbitrary gradient from the desired level of output back to the input.}\label{fig. 5}
\end{figure*}

\begin{figure*}[t!]
\centering
\includegraphics[width=11.4cm]{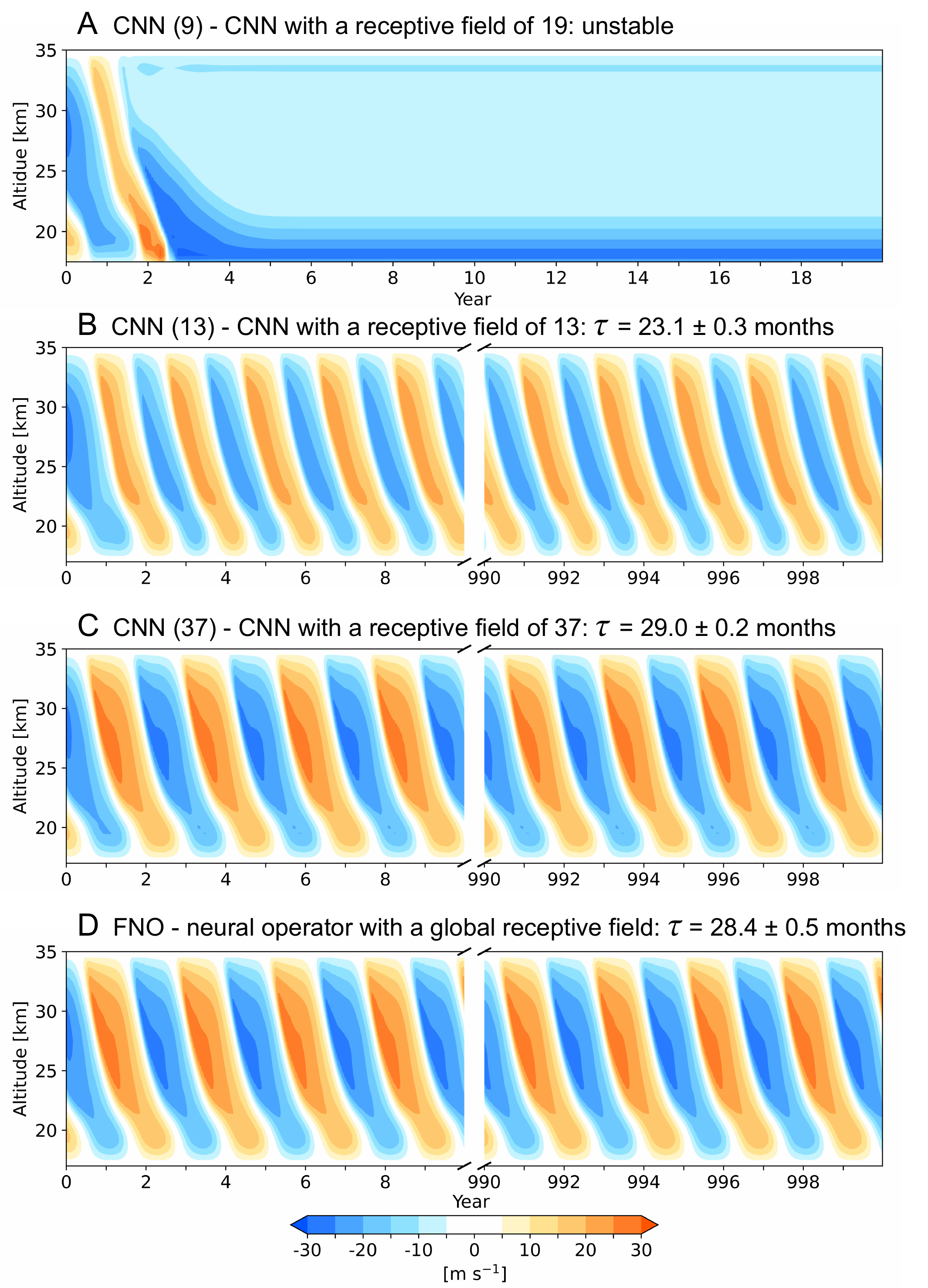}
\caption{Emulation of the 1D QBO model with (\textit{A-C}) CNNs having different RFs, as indicated, and (\textit{D}) the FNO. The emulator evolves the zonal wind autoregressively from \(u_t\) to \(u_{t+\Delta t}\), where \(\Delta t\) is 4 days (the dynamical time step is 1 day).}\label{fig. 6}
\end{figure*}


\paragraph{B. Receptive field (RF).} The RF of an NN is the region in the input domain that influences the prediction of a specific region of the output, meaning information outside the RF is not used for that prediction. For CNNs, the RF size depends on their architecture and can be mathematically calculated (Eq.~\ref{eq:5}). The RF can be increased by adding layers, increasing the kernel size, or increasing the dilation or stride (\textit{Materials and Methods C}).

Figure~\ref{fig. 2}\textit{A} provides a schematic of the RF for an output in the middle of the vertical domain for CNN(19), whose online performance is shown in Fig.~\ref{fig. 1}\textit{B}. This four-layer CNN has an RF of 19, meaning each output at the final layer can see only a region of 19 levels of the input. For a 1D model with a vertical resolution of 500 m, this covers almost half of the input domain, as shown in Fig.~\ref{fig. 2}\textit{A}. In other words, the CNN can estimate the GW drag at a given level based on 9.5 km of the zonal wind profile around that level.

In contrast, the CNN in Fig.~\ref{fig. 1}\textit{C}, which leads to stable and accurate simulations, has an RF of 55, allowing it to see the entire input domain. This is also true for the GW parameterization based on MLP and FNO, which theoretically have a global RF. This suggests that achieving a stable NN-based GW parameterization may require the NN to see the entire input domain. Specifically, a CNN-based GW parameterization may require an RF large enough to encompass the entire vertical profile of zonal wind to achieve stable simulations. This hypothesis aligns with the vertical coupling and non-local dynamics of GW propagation and dissipation \cite{campbell2005constraints}, which are further investigated in the following sections.

To test this hypothesis, we train 180 CNNs with varying RFs for several different vertical resolutions of the 1D model. The model's vertical domain is constant, spanning from 17 to 35 km. As we increase the model's resolution, the number of levels increases. For example, a resolution of 1500 m results in 11 levels, while a resolution of 100 m results in 179 levels. We run the 1D model at 14 different resolutions between these two extremes and train CNNs with varying RFs for each resolution. All these CNNs have roughly the same number of parameters (15,000) and demonstrate excellent offline performance (\(R^2 \approx 0.99\)). However, their RFs, which range from 5 to 241, were adjusted by changing the number of layers, the kernel size, or the dilation. The RFs of these CNNs for different 1D model resolutions are shown in Fig.~\ref{fig. 2}\textit{B}.

The striking finding is that for a CNN to remain stable and accurate, its RF must exceed the number of model levels. The key takeaway is that, in our testbed, the RF can serve as a quantitative metric, calculated based on the CNN's architecture, to predict instability a-priori. However, this is not the case for the RMSE, a commonly used metric, as shown in Fig.~\ref{fig. 2}\textit{C}. While CNNs tend to become more stable as the RMSE decreases, online stability cannot be predicted solely based on RMSE, consistent with previous studies. Interestingly, there are several examples where CNNs with nearly identical RMSE values can be either stable or unstable. This underscores the predictive power of the RF in our testbed. The offline and online performances of these CNNs, as well as details on their architectures for a few selected resolutions as examples, are provided in \textit{SI Appendix Table S1}.

Here, an NN is considered stable if the standard deviation of the QBO period, based on a 1000-year simulation, is not more than 10\% different from 0.7 months, which is the standard deviation of our true QBO period. This ensures timely and accurate transitions between QBO phases. In the following sections, we explain why instability primarily affects the QBO period and is most pronounced during transition phases.

Additionally, for a CNN with a fixed RF of 55, the offline error increases with higher resolution and more model levels, especially in the top 2 km (\textit{SI Appendix, Fig. S1}). This trend is not observed for an FNO, where the error increase is less pronounced and not dominated by the highest levels (\textit{SI Appendix, Fig. S1}). This difference arises from how FNOs and CNNs handle spatial information. CNNs operate on grid indices \cite{Goodfellow-et-al-2016}, applying convolutional filters without considering physical distances, making them agnostic to actual physical lengths. In contrast, FNOs use the Fourier transform to work in the frequency domain, capturing global spatial relationships and physical lengths \cite{li2020fourier}. This allows FNOs to model problems more accurately where the physical dimensions and continuous space are critical, making them better suited for applications that require an understanding of non-local dynamics \cite{azizzadenesheli2024neural}.

Next, we will demonstrate that the differences in stability between the CNNs, despite their similar size and offline performance (Fig.~\ref{fig. 2}\textit{B}), can be attributed to their varying capacities to learn the non-local dynamics of GW propagation and dissipation. Although we present results for the 500 m resolution and primarily focus on CNN(19) and CNN(55), the same conclusions apply to other vertical resolutions and CNN architectures.

\paragraph{C. Non-locality of GW dynamics.}

GWs can be triggered by various mechanisms, such as convection, and then propagate both horizontally and vertically, affecting distant regions by transporting energy and momentum through the atmosphere \cite{fritts2003gravity}. In the 1D framework, GW drag at any height depends on the cumulative wind profile below, down to the GW source level, as implied by the integral operator in Eq.~\ref{eq:2}. This non-local and vertical coupling is integrated into most physics-based GW parameterization schemes and is crucial for generating a spontaneous QBO \cite{campbell2005constraints}. However, these schemes do not currently account for horizontal non-locality, leading to significant errors \cite{kruse2022observed, kim2024crucial}. Novel data-driven GW parameterizations aim to address these shortcomings \cite{gupta2024machine}. While our focus is on the 1D model and vertical coupling, our findings can also apply to horizontal non-locality.

The vertical coupling is also evident in a correlation matrix between GW drag and zonal wind for the 1D model, as shown in Fig.~\ref{fig. 3}\textit{A}. For example, GW drag at the 34 km level (\(x\)-axis) is anticorrelated with a deep vertical profile of zonal wind (\(y\)-axis). Such a non-local correlation is also evident in the observed QBO based on ERA5 reanalysis data (Fig.~\ref{fig. 3}\textit{B}). It should also be mentioned that local dynamics, particularly the zonal wind shear, are crucial for GW breaking (Figs.~\ref{fig. 3}\textit{C} and \textit{D}), and GW drag usually maximizes near strong shear zones. Thus, both local and non-local dynamics are important and need to be captured for a stable and accurate data-driven GW parameterization \cite{pahlavan2024explainable}.

With this in mind, we examine the ability of different NN architectures to learn this long-range non-locality. Figure~\ref{fig. 4}\textit{A} shows the vertical profile of offline error for CNN(19) compared to three other NNs. While overall errors are relatively small and similar across all NNs, the error in CNN(19) notably increases with height, especially above 24 km. Figure~\ref{fig. 4}\textit{B} shows the difference between the correlation matrix of the true QBO and that generated by CNN(19). The blue color indicates a weaker correlation between the GW drag at the highest levels and the zonal wind at the lowest levels in CNN(19) compared to the true QBO. This type of error is not observed in CNN(55) (Fig.~\ref{fig. 4}\textit{C}), FNO (Fig.~\ref{fig. 4}\textit{D}), or MLP (not shown). Notably, increasing CNN parameters from 15,000 to 60,000 by adding more filters per layer, without changing the RF, does not affect this result (\textit{SI Appendix, Fig. S2}). Therefore, the error increase with height in CNN(19) can be attributed to its inability to learn the non-local relationship due to its smaller RF. This also explains the pronounced error increase at the top of the model for a CNN with a fixed RF when the resolution of the 1D model, and thus the number of levels, increases (\textit{SI Appendix, Fig. S1}).

\paragraph{D. Effective receptive field (ERF).}

RF can be analytically derived and depends solely on the architecture of CNNs, remaining constant during training and independent of the weights and input data. For FNO and MLP, the RF theoretically covers the entire input domain. Conversely, the effective receptive field (ERF) can be numerically calculated for any NN architecture and depends on the weights and input data. The ERF quantifies how much a specific output is affected by a slight change in the input, identifying which part of the RF is most influential. In our case, it measures the extent to which the GW drag at a specific level is affected by a slight change in the zonal wind (i.e., \(\partial{G}/\partial{u}\)). Calculating the ERF involves backpropagating a signal from the output layer to the input layer (\textit{Materials and Methods D}), similar to the layer-wise relevance propagation (LRP) method \cite{montavon2018methods}, which has been recognized as an informative NN visualization method for climate applications \cite{toms2020physically, ebert2020evaluation}.

The ERF of the different NNs for GW drag at four different levels is shown in Fig.~\ref{fig. 5}. Figure~\ref{fig. 5}\textit{A} illustrates that for GW drag at the 18 km level, the zonal wind profiles at the 18 km level and nearby levels are the most influential. The impact sharply decreases above 19 km. Despite some differences, the overall shape of the profiles for different architectures is similar.

For the GW drag at the 34 km level (Fig.~\ref{fig. 5}\textit{B}), while the nearby levels are again most influential, the NNs, except for CNN(19), use information from the lowermost levels of the zonal wind profile to calculate the drag, as expected from Figs.~\ref{fig. 3}\textit{A} and \textit{C}. However, the ERF for CNN(19) below the 27.5 km level is effectively zero due to its smaller RF, resulting in a markedly different ERF shape compared to the other three NNs. This inability of CNN(19) to learn the non-local relation between GW drag at the highest levels and zonal wind at the lowest levels renders the simulations unstable.

For GW drag at the middle levels (Figs.~\ref{fig. 5}\textit{C} and \textit{D}), the most influential levels are also the surrounding ones, with a sharp decrease above, especially for CNN(55) and FNO, but remaining non-negligible for all levels below. These patterns align well with the expected dynamics of GW propagation and dissipation, as well as the correlation matrices in Figs.~\ref{fig. 3}\textit{A} and \textit{C}. The GW drag at a specific level is influenced by the local zonal wind, with waves propagating upward more slowly when the wind is near their phase speed, leading to enhanced dissipation near critical levels and strong shear zones, highlighting the essential role of local dynamics (Fig.~\ref{fig. 3}\textit{C}). Additionally, the cumulative wind profile below a given level significantly impacts the GW drag, highlighting the importance of non-local dynamics (Fig.~\ref{fig. 3}\textit{A}). Importantly, the limitation of the smaller RF in CNN(19) becomes evident again in Figs.~\ref{fig. 5}\textit{C} and \textit{D}. 

These results further explain the increase in error with height for CNN(19) (Fig.~\ref{fig. 4}\textit{A}) and its inability to learn the non-local correlation between GW drag at the highest levels and zonal wind at the lowest levels (Fig.~\ref{fig. 4}\textit{B}). These also explain the behavior seen in \textit{Fig. S1} (\textit{SI Appendix}): as the resolution of the 1D model and the number of levels increase, the percentage of the input domain covered by a fixed RF decreases. Consequently, the error increases, especially for estimating the GW drag at the top levels, which requires information from the zonal wind at the lowermost levels.

\paragraph{E. Spatiotemporal variability of error.}

One notable observation is that instability is most pronounced during transition phases, as shown in Fig.~\ref{fig. 1}\textit{B}, and primarily affects the QBO period (\textit{SI Appendix Table S1}). To better understand this behavior, we investigate the temporal variability of the error by plotting it against the QBO phase, calculated using principal component analysis (\textit{SI Appendix Fig. S3}). As expected, the error is largest for CNN(19) and relatively negligible for the other NNs, particularly the FNO. However, CNN(19) exhibits an interesting spatiotemporal error pattern: it maximizes not only at the highest levels but also during the QBO transition phases. This suggests that the error significantly impacts the initiation of new QBO phases, leading to unrealistic variability in the QBO period compared to the true QBO in the 1D model. Therefore, we use the standard deviation of the period as the measure of stability.

\paragraph{F. Data-driven emulator of the 1D model.}

So far, we have demonstrated the importance of learning non-local dynamics for achieving stable and accurate data-driven parameterizations in the 1D testbed. However, the climate system is inherently non-local due to the propagation of waves and teleconnections, which create spatial patterns linking distant weather and climate phenomena. Therefore, it is reasonable to expect that learning non-local dynamics is also critical for building stable and accurate data-driven emulators of the entire climate system. We confirm this by emulating the 1D QBO model, where the emulator evolves the zonal wind autoregressively from \(u_t\) (input to the NN) to \(u_{t+\Delta t}\) (output of the NN), with \(\Delta t\) being 4 days (the dynamical time step for the 1D model is 1 day).

Figure \ref{fig. 6}\textit{A} demonstrates that a CNN with an RF of 9 becomes unstable very quickly despite its excellent offline performance (\(R^2 \approx 0.99\) between the true and CNN-predicted \(u_{t+\Delta t}\)). Increasing the CNN's RF to 13 makes the emulator somewhat stable but still inaccurate (Fig.\ref{fig. 6} \textit{B}), with a mean period significantly shorter than the true period (23.1 vs. 28.7 months). However, when the RF is large enough (Fig.\ref{fig. 6} \textit{C}) or when an FNO is used (Fig.\ref{fig. 6}\textit {D}), the emulator becomes stable and accurate. These NNs have the same number of parameters as those used previously (15,000). These results further emphasize the critical importance of learning non-local dynamics for achieving stable and accurate data-driven algorithms.

It is worth mentioning that the emulated zonal wind profiles shown in Fig.~\ref{fig. 6} are too smooth compared to the true QBO (Fig.~\ref{fig. 1}\textit{A}), missing the stochastic variability and resulting in a smaller standard deviation of the period. This issue of smoothing/blurring is due to a phenomenon called the spectral bias of NNs \cite{chattopadhyay2023long} and is common in the emulation of multi-scale dynamical systems such as weather and climate \cite{bonavita2024some}. However, a more detailed investigation of this problem is left for future study.

\section*{Discussion}
\label{sec:Discussion}

Data-driven, particularly ML-based, parameterizations present a promising alternative to traditional physics-based methods, with the potential to reduce uncertainties in climate modeling. However, despite their high offline accuracy, these parameterizations often cause instabilities when integrated into numerical solvers, typically resolved through trial and error and ad-hoc remedies. Such instabilities also affect data-driven emulators of weather and climate. By integrating climate physics and ML theory, this work identifies one source of these instabilities and presents a rigorous approach for addressing these instabilities. This also further highlights the necessity of such an integration of physics and ML to develop rigorous and systematic methods for data-driven weather/climate modeling.

Using a 1D model of the QBO and GW parameterization as a testbed, we demonstrated that $R^2$ and RMSE, common metrics for offline performance, cannot predict instability a-priori. We showed that this instability arises from NNs' inability to learn non-local dynamics. Additionally, we used RF, a quantitative metric based on ML theory, to predict instability before coupling NN-based GW parameterizations to the 1D model. We explored three NN architectures (CNN, FNO, and MLP), representing broad NN classes, and highlighted the usefulness of the effective RF (ERF), which can be calculated for any architecture to quantify the degree to which non-locality is captured.

While this study mainly focused on GW parameterization and the importance of learning non-local dynamics in the vertical direction, the findings broadly apply to other data-driven SGS models, especially those inherently non-local due to their large vertical, horizontal, or temporal extents. For instance, convective processes involve the vertical transport of moisture and momentum, making them non-local due to the large vertical extent of convective clouds \cite{emanuel1994atmospheric}. Another example is turbulent mixing in the planetary boundary layer, which redistributes heat and moisture over large vertical and horizontal scales \cite{stull2012introduction}. Radiative transfer processes can also occur over large spatial scales \cite{liou2002introduction}. Several studies have highlighted the significance of including non-locality, either in space or time, to improve SGS modeling \cite{palmer2001nonlinear, bolton2019applications, han2020moist, gupta2024machine}. For instance, Wang et al. \cite{wang2022non} trained NNs using non-local inputs spanning over 3 × 3 columns around the target column, finding that including these non-local inputs significantly improves the offline prediction of various SGS processes, particularly for SGS momentum transport.

We have further demonstrated the importance and necessity of learning non-local dynamics in building stable and accurate fully data-driven emulators. It is crucial to recognize that non-locality is ubiquitous in the climate system due to propagating features like GWs and Rossby waves that redistribute energy and momentum horizontally and vertically, affecting weather patterns and climate. Additionally, teleconnections from variability modes like the El Niño-Southern Oscillation (ENSO), Madden-Julian Oscillation (MJO), and the QBO create spatial patterns linking distant weather and climate phenomena. With the rapid rise of weather and climate emulators \cite{ben2024rise, lai2024machine}, where instability is also a common issue \cite{chattopadhyay2023long}, our findings highlight the need to consider non-localities, both in the vertical and horizontal directions, in architectural design to build more accurate and stable emulators. More broadly, this work emphasizes the need for rigorous methods/criteria for applications of NNs to weather/climate modeling for both hybrid and fully data-driven approaches.

\section*{Materials and Method}
\label{sec:Materials}

\paragraph{A. 1D-QBO model.}
The 1D-QBO model represents a one-dimensional prototype of the tropical stratosphere \cite{holton1972updated, plumb1977interaction}. It includes a source of parameterized waves at its lower boundary, serving as a minimal configuration to capture the fundamental dynamics of wave-mean flow interaction. In this study, the 1D-QBO is structured as a forced advection-diffusion model:
\begin{equation} \label{eq:1}
\frac{\partial{u}}{\partial{t}} + \omega \frac{\partial{u}}{\partial{z}} - \kappa \frac{\partial^2{u}}{\partial{z}^2} = G(u) + \eta(t),
\end{equation}

\noindent with zonal wind $u (t, z)$ as a function of time $t$ and height $z$, upwelling $\omega = 0.1$ mm s$^{-1}$ , diffusivity $\kappa= 0.4 $ m$^2$ s$^{-1}$, and GW forcing $G$. The term $\eta$ is a stochastic forcing, which represents missing physics within the simple 1D model \cite{pahlavan2024explainable}.

Following \cite{plumb1977interaction}, the 1D-QBO model is driven by two GWs of equal and opposite phase speed with vertical group velocity $c_{gz} = k(u - c_n)^2/N$ for the discrete wave $n$, with buoyancy frequency $N = 2.16 \times 10^{-2}$ s$^{-1}$, wavenumber $k = 2\pi/40000$ km$^{-1}$, and phase speed $(c_1, c_2) = (-30, +30)$ m s$^{-1}$. The model domain is from $z_L = 17$ km to $z_T = 35 $ km. Subscript $L$ refers to the lower boundary, and subscript $T$ to the upper boundary. Assuming a constant wave dissipation rate $\alpha = 1.2 \times 10^{-6}$ s$^{-1}$, the wave momentum flux as a function of height for a wave with source momentum flux $F_L = 6.325 \times 10^{-3}$ m$^2$ s$^{-2}$ is:

\begin{equation} \label{eq:2}
F_n(u,z) = F_L \, \mathrm{sgn}(c_n)\exp \, \biggl\{- \int_{z_L}^{z}\frac{\alpha}{c_{gz}}dz^{'}\biggl\}.
\end{equation}
Note that this integral over height introduces non-locality in the relationship between $u$ and $G$, consistent with the dynamics of GWs and their breaking and dissipation in the real atmosphere. 

The GW drag is proportional to the divergence of momentum flux summed across all waves:

\begin{equation} \label{eq:3}
G(u,z) = -\frac{\rho_L}{\rho_0}\sum_{n}\frac{\partial F_n}{\partial z},
\end{equation}

\noindent with density $\rho_0 = \rho_L \mathrm{e}^{{(z-z_L)}/H_\rho}$ with $\rho_L=0.1$ kg m$^{-1}$, and $H_\rho = 6$ km.

We run this model at 16 different vertical resolutions, ranging from 100 m to 1500 m, which yields a varying number of vertical levels between 179 and 11. At the 500 m resolution (35 levels), this setup yields an oscillation with a period ($\tau$) of 28.7$\pm$0.7 months, and an amplitude ($\sigma$) of 20.1$\pm$0.3 m s$^{-1}$ at the 25 km altitude (Figure~\ref{fig. 1}\textit{A}). The period and amplitude of the QBO at a few other resolutions are shown in \textit{SI Appendix, Table S1}.

\paragraph{B. NN-based GW parameterization.}

We represent the GW drag using three different NN architectures: CNN, FNO, and MLP, each representing a broad class of algorithms. For each class, we performed extensive hyperparameter optimization (HPO) to identify the optimal hyperparameters, including learning rate, batch size, loss function, activation function, and optimizer. Details of the architectures of the NNs are provided in the shared code (\textit{E. Data, Materials, and Software Availability}). The NNs are trained offline by minimizing the following loss:

\begin{equation} \label{eq:4}
\mathcal{L}\, = \frac{1}{m} \sum_{i=1}^{m} \,\,  \bigl\| G(u_i) - G_{\mathrm{NN}}(u_i, \theta) \bigl\|_2^2
\end{equation}

\noindent where $m$ is the number of training samples, $\theta$ represent the learnable parameters of the NN, and $\|.\|_2$ is the $L_2$ norm.

\paragraph{C. Receptive field (RF).}

We aim to determine the size of the input region that influences the output of an NN. This is determined by the RF. For generic CNNs, the RF can be mathematically derived \cite{araujo2019computing}:

\begin{equation} \label{eq:5}
\mathrm{RF} = \sum_{l=1}^{D} \Bigl(\delta_l(\kappa_l - 1) \prod_{i=1}^{l-1} s(i) \Bigl) + 1
\end{equation}

\noindent with the number of layers $D$, dilation size $\delta$, kernel size $\kappa$, and stride size $s$. This expression is intuitive when considering some special cases. For instance, if all kernels are of size 1, the RF will naturally be size 1. If all strides and dilations are 1, the RF is simply the sum of $(\kappa_l - 1)$ over all layers, plus 1, which is straightforward.

\paragraph{D. Effective receptive field (ERF).}

While the RF is an inherent architectural characteristic of a CNN, the ERF can be numerically calculated for any NN architecture. The ERF assesses the relative importance of each input in estimating a specific output \cite{luo2016understanding}. We use the partial derivative \(\partial{G}/\partial{u}\) to measure the impact, which indicates how much \(G\) at a specific level changes as \(u\) varies slightly at all levels. Therefore, this derivative serves as a natural measure of the importance of the input \(u\) at different levels for estimating the output \(G\) at a given level. Because this measure depends on both the NN's weights and the details of the input (Fig.~\ref{fig. 5}), our results are presented as expectations over the input distribution, i.e., averaged over the entire test set. The partial derivatives can be computed by back-propagating an arbitrary gradient from the output to the input, similar to the standard process of propagating the error gradient with respect to a loss function \cite{luo2016understanding}. The gradient is set to zero for all outputs except the region of interest, so the ERF reveals the sensitivity of this specific output to changes in the input. For further details on this analysis, refer to \cite{luo2016understanding}.

\paragraph{E. Data, Materials, and Software Availability.}

The code for the 1D-QBO model is available on \href{https://github.com/DataWaveProject/qbo1d}{GitHub}. You can also access the code for training the neural networks via this \href{https://github.com/HamidPahlavan/Nonlocality}{link}.

\paragraph{ACKNOWLEDGMENTS.}

We are grateful to Moein Darman for providing the one-dimensional code for the FNO. This work is supported by grants from the NSF OAC CSSI program (2005123 and 2004512), Schmidt Sciences, LLC (to P.H. and J.A.), NSF award 2046309 (to P.H.), and the Office of Naval Research (ONR) Young Investigator Award N00014-20-1-2722 (to P.H.). Computational resources are provided by NCAR's CISL (allocation URIC0009) and NSF XSEDE (allocation ATM170020).

\bibliographystyle{unsrt}  

\bibliography{template}

\break

\section*{Supporting Information}
\label{sec:SI}

\vfill \break

\setcounter{figure}{0}

\makeatletter 
\renewcommand{\thefigure}{S\@arabic\c@figure}
\makeatother

\begin{figure}
\centering
\includegraphics[width=11.4cm]{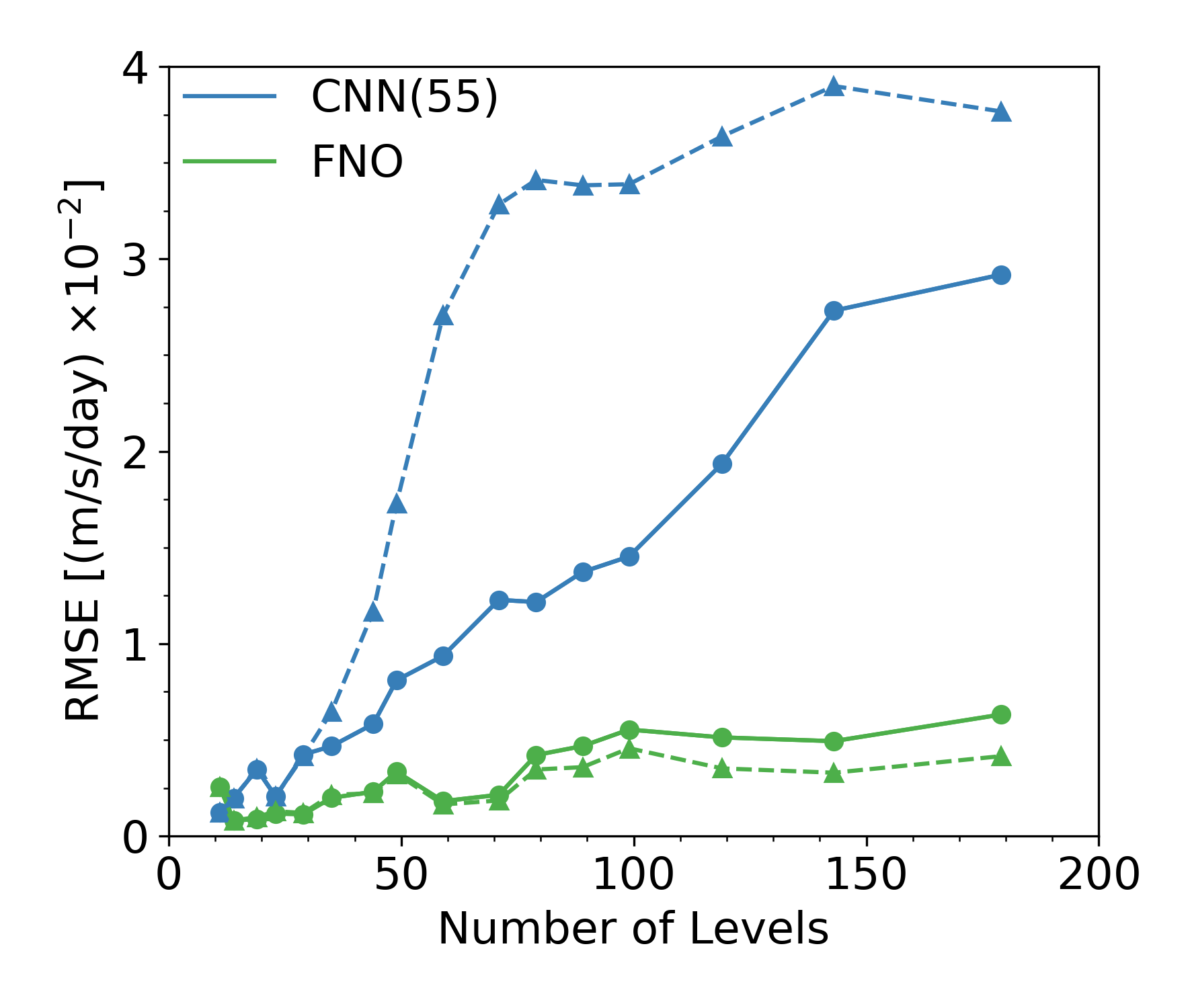}
\caption{Root-mean-square-error (RMSE) for a CNN with a fixed RF of 55, and for an FNO, across different resolutions of the 1D model. The solid line represents the average error over all levels, while the dashed lines represent the average error over the top 2 km of the model.}\label{fig. S1}
\end{figure}

\begin{figure}
\centering
\includegraphics[width=17.8cm]{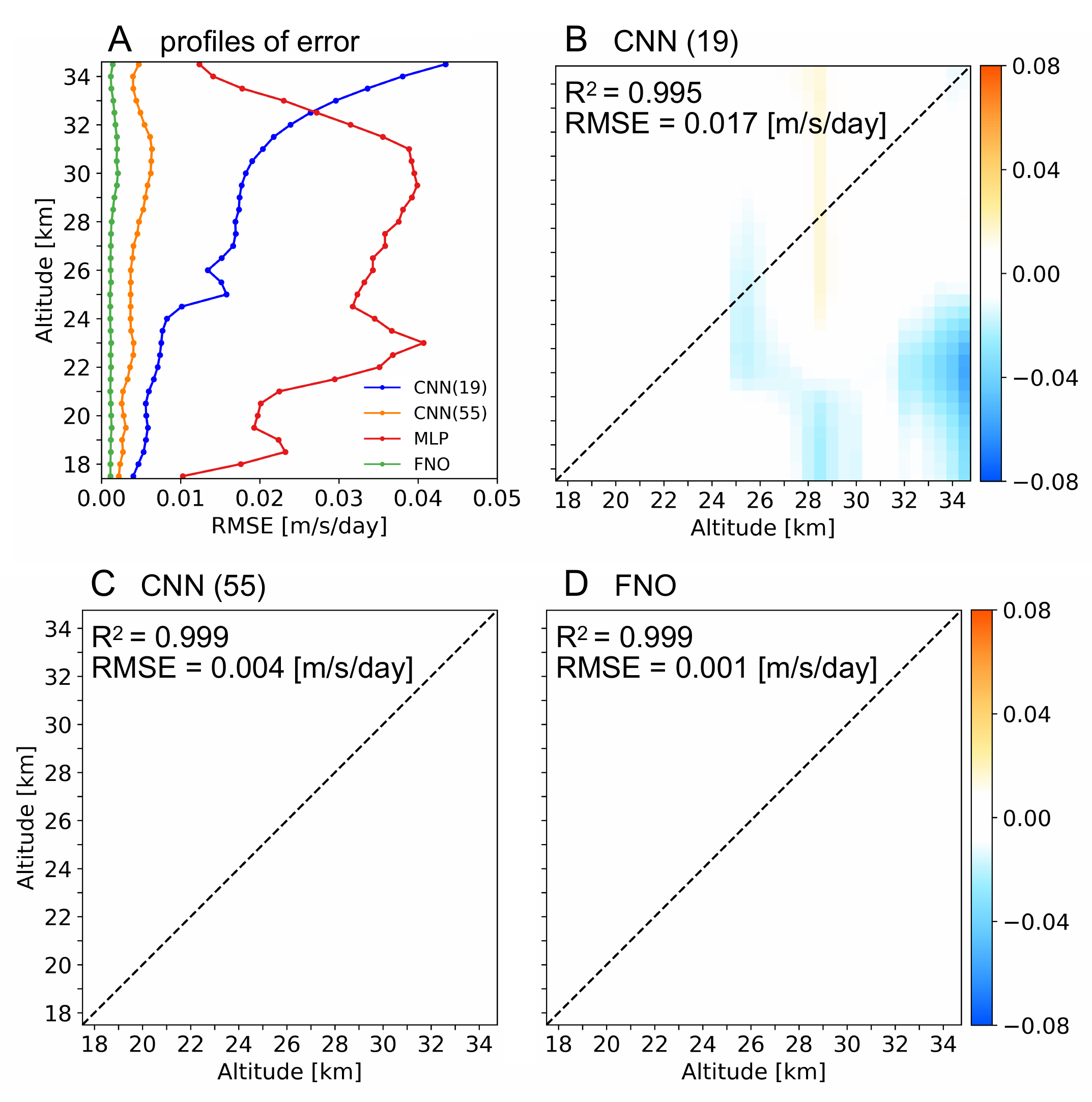}
\caption{As in Fig. 4, but here each NN has 60,000 parameters, unlike the 15,000 in Fig. 4. While the overall behavior of CNN(19) and CNN(55) remains the same due to their unchanged RF, the MLP performs worse, likely due to overfitting.}\label{fig. S2}
\end{figure}

\begin{figure}
\centering
\includegraphics[width=17.8cm]{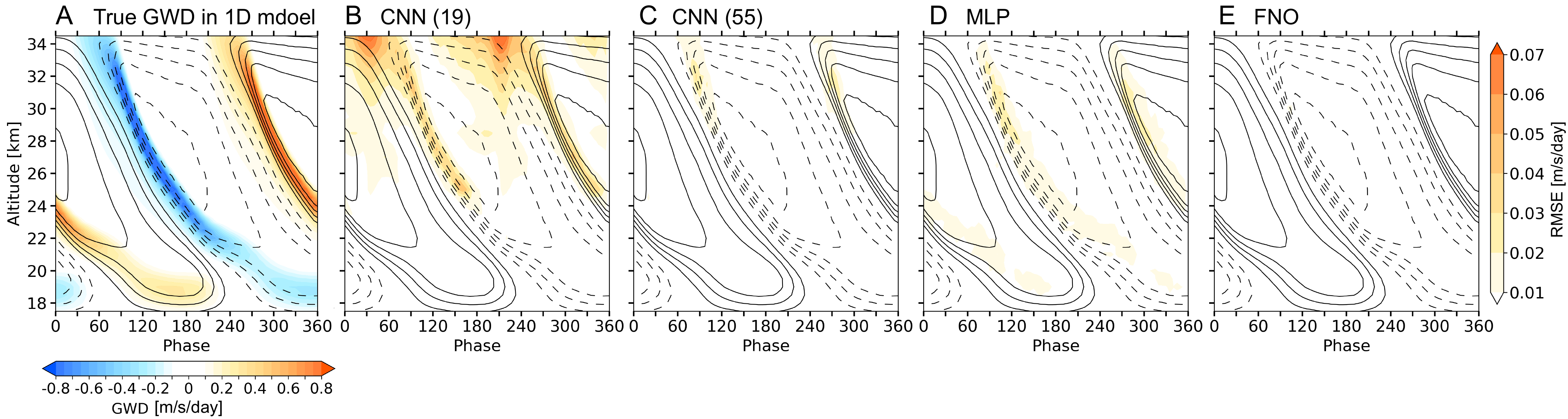}
\caption{Variability of error as a function of height and phase of the QBO. (\textit{A}) The true GWD in the 1D model, and (\textit{B-E}) the offline error of various NN architectures, as indicated. Contours show the zonal wind. The contour interval is 5 m s$^{-1}$, westerlies are solid, easterlies are dashed, and the zero contours are omitted.}\label{fig. S3}
\end{figure}

\break

\makeatletter 
\renewcommand{\thetable}{S\@arabic\c@table}
\makeatother

\begin{table}
\centering
\caption{Details of the architecture and performance of the CNNs shown in Figs. 2\textit{B} and \textit{C} for four selected resolutions. Offline performance is evaluated based on root-mean-square-error (RMSE) and pattern correlation (\boldmath\(R^2\)), while online performance is assessed based on the mean amplitude (\boldmath$\sigma_m$) and its standard deviation (\boldmath$\sigma_s$), as well as the mean period (\boldmath$\tau_m$) and its standard deviation (\boldmath$\tau_s$) of the QBO. True values for these QBO characteristics, based on a 1000-year simulation, are shown in parentheses for each resolution of the 1D model, with minor differences from varying resolutions. Blue indicates a stable QBO simulation, while red denotes an unstable model with a \boldmath$\tau_s$ over 10\% different from the true QBO.}

\begin{tabular} {p{2.47cm} |p{0.75cm}|p{0.75cm}|p{0.75cm}|p{0.75cm}|p{0.75cm}|p{0.75cm}|p{0.75cm}|p{0.75cm}|p{0.75cm}|p{0.75cm}|p{0.75cm}|p{0.75cm}}
\hline\hline

500 m (35 levels) & \cellcolor{red} & \cellcolor{red}  & \cellcolor{blue}  & \cellcolor{blue} & \cellcolor{red} & \cellcolor{blue} & \cellcolor{blue} & \cellcolor{blue}  & \cellcolor{blue} & \cellcolor{blue} & \cellcolor{blue} & \cellcolor{blue}  \\
 \hline
\scriptsize{Number of layers} & 4	&4	&4	&4	&6	&8	&9	&10	&12	&4	&4	&4 \\
\scriptsize{Kernel size} & 7	&9	&15	&19	&7	&7	&7	&7	&7	&7	&7	&7 \\
\scriptsize{Number of channels} & 33	&28	&22	&19	&23	&19	&17	&16	&14	&33	&33&	33 \\
\scriptsize{Dilation} & 1 &1	&1	&1	&1	&1	&1	&1	&1	&2	&3	&4 \\
\scriptsize{Number of parameters} & 15808	&14701	&15247	&14498	&15250	&15562	&14536	&14705	&14071	&15808	&15808	&15808 \\
\scriptsize{Receptive field}  & 19	&25	&43	&55	&31	&43	&49	&55	&67	&37	&55&	73 \\
 \hline
RMSE ($\times10^2$) & 1.64	&1.37	&0.53	&0.48&	1.26	&0.73	&0.58	&0.40	&0.51&	1.54	&2.03	&2.94 \\
$R^2$ & 0.996	&0.997	&1.000	&1.000	&0.997	&0.999	&0.999	&1.000	&1.000	&0.996	&0.993	&0.986 \\
 \hline
$\sigma_{m}$  (20.1 m/s) & 19.3	&19.8	&20	&20	&20	&20.1	&20	&20.1	&20.1	&20	&20.1	&20.1 \\
$\sigma_{s}$  (0.3 m/s) & 1.6 &0.4	&0.3& 0.3&	0.3&	0.3&	0.3&	0.3&	0.3&	0.3&	0.3&	0.3 \\
$\tau_{m}$  (28.7 months) & 30.2 &28.5	&28.8	&28.8	&28.7	&28.8	&28.8	&28.8	&28.8	&28.8	&28.8&	28.9 \\
$\tau_{s}$  (0.72 months) & 11.25 &1.69	&0.69	&0.72	&1.06	&0.68	&0.69	&0.68	&0.67	&0.71	&0.69	&0.75 \\
\bottomrule

400 m (44 levels) & \cellcolor{red} & \cellcolor{red}  & \cellcolor{blue}  & \cellcolor{blue} & \cellcolor{blue} & \cellcolor{red} & \cellcolor{red} & \cellcolor{red}  & \cellcolor{blue} & \cellcolor{blue} & \cellcolor{red} & \cellcolor{blue}  \\
 \hline
\scriptsize{Number of layers} & 5 & 5 & 5 & 5 & 5 & 6 & 7 & 8 & 9 & 10 & 4 & 4 \\
\scriptsize{Kernel size} & 5 & 9 & 13 & 17 & 21 & 7 & 7 & 7 & 7 & 7 & 7 & 9 \\
\scriptsize{Number of channels} & 31 & 23 & 19 & 17 & 15 & 23 & 21 & 19 & 17 & 16 & 33 & 28 \\
\scriptsize{Dilation} & 1 & 1 & 1 & 1 & 1 & 1 & 1 & 1 & 1 & 1 & 2 & 2 \\
\scriptsize{Number of parameters} & 14850 & 14790 & 14650 & 15386 & 14866 & 15250 & 15856 & 15562 & 14536 & 14705 & 15808 & 14701 \\
\scriptsize{Receptive field}  & 17 & 33 & 49 & 65 & 81 & 31 & 37 & 43 & 49 & 55 & 37 & 49 \\
 \hline
RMSE ($\times10^2$) & 2.38 & 1.35 & 0.82 & 0.50 & 0.59 & 1.40 & 1.31 & 1.21 & 1.04 & 0.79 & 1.76 & 1.46 \\
$R^2$ & 0.992 & 0.997 & 0.999 & 1.000 & 0.999 & 0.997 & 0.997 & 0.998 & 0.998 & 0.999 & 0.995 & 0.997 \\
 \hline
$\sigma_{m}$  (20.1 m/s) & 19.9 & 20.0 & 20.1 & 20.1 & 20.1 & 19.8 & 20.0 & 20.1 & 20.1 & 20.1 & 20.1 & 20.1 \\
$\sigma_{s}$  (0.3 m/s) & 0.5 & 0.3 & 0.3 & 0.3 & 0.3 & 0.9 & 0.3 & 0.3 & 0.3 & 0.3 & 0.3 & 0.3 \\
$\tau_{m}$  (28.6 months) & 27.3 & 28.3 & 28.7 & 28.7 & 28.7 & 28.3 & 28.5 & 28.6 & 28.6 & 28.7 & 28.7 & 28.7 \\
$\tau_{s}$  (0.72 months) & 2.96 & 1.39 & 0.69 & 0.71 & 0.72 & 4.17 & 1.27 & 0.86 & 0.74 & 0.64 & 0.85 & 0.65 \\
\bottomrule

360 m (49 levels) & \cellcolor{red} & \cellcolor{red}  & \cellcolor{blue}  & \cellcolor{blue} & \cellcolor{blue} & \cellcolor{red} & \cellcolor{blue} & \cellcolor{blue}  & \cellcolor{blue} & \cellcolor{blue} & \cellcolor{red} & \cellcolor{blue}  \\
 \hline
\scriptsize{Number of layers} & 5     & 5     & 5     & 5     & 5     & 6     & 7     & 8     & 9     & 10    & 4     & 5 \\
\scriptsize{Kernel size} & 5     & 9     & 13    & 17    & 21    & 9     & 9     & 9     & 9     & 9     & 7     & 7 \\
\scriptsize{Number of channels} & 31    & 23    & 19    & 17    & 15    & 20    & 18    & 16    & 15    & 14    & 32    & 26 \\
\scriptsize{Dilation} & 1     & 1     & 1     & 1     & 1     & 1     & 1     & 1     & 1     & 1     & 2     & 2 \\
\scriptsize{Number of parameters} & 14850 & 14790 & 14650 & 15386 & 14866 & 14861 & 15013 & 14225 & 14566 & 14491 & 14881 & 14665 \\
\scriptsize{Receptive field}  & 17    & 33    & 49    & 65    & 81    & 41    & 49    & 57    & 65    & 73    & 37    & 49  \\
 \hline
RMSE ($\times10^2$) & 2.64  & 1.42  & 1.02  & 0.69  & 0.36  & 1.29  & 1.10  & 0.85  & 0.69  & 0.49  & 1.80  & 1.57 \\
R$^2$ & 0.990 & 0.997 & 0.998 & 0.999 & 1.000 & 0.998 & 0.998 & 0.999 & 0.999 & 1.000 & 0.995 & 0.996 \\
 \hline
$\sigma_{m}$  (20.1 m/s) & 19.9  & 19.9  & 20.0  & 20.0  & 20.0  & 20    & 20    & 20    & 20.1  & 20.1  & 20    & 20 \\
$\sigma_{s}$  (0.3 m/s) & 0.3   & 0.4   & 0.3   & 0.3   & 0.3   & 0.3   & 0.3   & 0.3   & 0.2   & 0.2   & 0.3   & 0.3 \\
$\tau_{m}$  (28.6 months) & 26.5  & 28.4  & 28.7  & 28.6  & 28.7  & 28.4  & 28.6  & 28.6  & 28.7  & 28.7  & 28.4  & 28.7 \\
$\tau_{s}$  (0.72 months) & 2.03  & 2.42  & 0.73  & 0.73  & 0.69  & 1.27  & 0.77  & 0.72  & 0.6   & 0.72  & 1.11  & 0.74  \\
\bottomrule

300 m (59 levels) & \cellcolor{red} & \cellcolor{red}  & \cellcolor{red}  & \cellcolor{blue} & \cellcolor{blue} & \cellcolor{red} & \cellcolor{red} & \cellcolor{red}  & \cellcolor{blue} & \cellcolor{blue} & \cellcolor{red} & \cellcolor{blue}  \\
 \hline
\scriptsize{Number of layers} & 5	&5	&5	&5	&5	&6	&7	&8	&9	&10	&5	&5\\
\scriptsize{Kernel size} & 5	&9	&13	&17	&21	&9	&9	&9	&9	&9	&7	&9 \\
\scriptsize{Number of channels} & 31	&23	&19	&17	&15	&20	&18	&16	&15	&14	&26	&23 \\
\scriptsize{Dilation} & 1&	1&	1&	1&	1&	1&	1&	1&	1&	1&	2&	2 \\
\scriptsize{Number of parameters} & 14850	&14790	&14650	&15386	&14866	&14861	&15013	&14225	&14566	&14491	&14665	&14790 \\
\scriptsize{Receptive field}  & 17&	33&	49&	65&	81&	41&	49&	57&	65&	73&	49&	65  \\
 \hline
RMSE ($\times10^2$) & 3.31	&1.69	&1.31	&1.27	&0.65	&1.48	&1.39	&1.27	&1.15	&0.99	&1.75	&1.39 \\
R$^2$ & 0.985&	0.996	&0.998	&0.998	&0.999	&0.997	&0.997	&0.998	&0.998&	0.999	&0.996&	0.997 \\
 \hline
$\sigma_{m}$  (20.1 m/s) & 15.9	&20.1&	20.1&	20.1&	20.1&	19.9&	19.9&	20.1&	20.1&	20.1&	20.1&	20.1 \\
$\sigma_{s}$  (0.3 m/s) & 2.7 &	0.6	&0.3	&0.3	&0.3	&0.7	&0.5	&0.3	&0.3	&0.3	&0.3	&0.3 \\
$\tau_{m}$  (28.6 months) & 39.4&	28&	28.2&	28.5&	28.6&	28&	28.3&	28.4&	28.5&	28.6&	28.4&	28.6 \\
$\tau_{s}$  (0.73 months) & 26.96&	3.4	&1.17	&0.77	&0.67	&3.24	&2.36	&1.08	&0.76	&0.68	&1.26	&0.73  \\
\bottomrule

\end{tabular}

\end{table}

\end{document}